\documentclass[conference]{IEEEtran}
\IEEEoverridecommandlockouts
\usepackage{cite}
\usepackage{amsmath,amssymb,amsfonts,systeme}
\usepackage{mathtools}
\usepackage[noend]{algorithmic}
\usepackage{graphicx}
\usepackage{textcomp}
\usepackage{amsmath}
\usepackage{xcolor}
\usepackage[ruled,vlined]{algorithm2e}
\def\BibTeX{{\rm B\kern-.05em{\sc i\kern-.025em b}\kern-.08em
    T\kern-.1667em\lower.7ex\hbox{E}\kern-.125emX}}

\makeatletter
\def\ps@IEEEtitlepagestyle{%
  \def\@oddfoot{\mycopyrightnotice}%
  \def\@evenfoot{}%
}
\def\mycopyrightnotice{%
  {\footnotesize \begin{tabular}[t]{@{}l@{}} © © 2022 IEEE. This paper appears in 2022 IEEE Wireless Communications and Networking Conference (WCNC 2022). Personal use of this material is \\ permitted. Permission from IEEE must be obtained for all other uses, in any current or future media, including reprinting/republishing this material for \\ advertising or promotional purposes, creating new collective works, for resale or redistribution to servers or lists, or reuse of any copyrighted component \\ of this work in other works.\end{tabular}}
  \gdef\mycopyrightnotice{}
}

\begin{document}

\title{IoTMonitor: A Hidden Markov Model-based Security System to Identify Crucial Attack Nodes in Trigger-action IoT Platforms \\
}

\author{\IEEEauthorblockN{Md Morshed Alam, Md Sajidul Islam Sajid, Weichao Wang, Jinpeng Wei}
\IEEEauthorblockA{Department of Software and Information Systems, University of North Carolina at Charlotte, Charlotte, USA \\
\{malam3, msajid, wwang22, jwei8\}@uncc.edu}
}


\maketitle

\begin{abstract}
With the emergence and fast development of trigger-action platforms in IoT settings, security vulnerabilities caused by the interactions among IoT devices become more prevalent. The event occurrence at one device triggers an action in another device, which may eventually contribute to the creation of a chain of events in a network. Adversaries exploit the chain effect to compromise IoT devices and trigger actions of interest remotely just by injecting malicious events into the chain. To address security vulnerabilities caused by trigger-action scenarios, existing research efforts focus on validation of the security properties of devices, or verification of the occurrence of certain events based on their physical fingerprints on a device. We propose IoTMonitor, a security analysis system that discerns the underlying chain of event occurrences with the highest probability by observing a chain of physical evidence collected by sensors. We use the Baum-Welch algorithm to estimate transition and emission probabilities and the Viterbi algorithm to discern the event sequence. We can then identify the crucial nodes in the trigger-action sequence whose compromise allows attackers to reach their final goals. The experiment results of our designed system upon the PEEVES datasets show that we can rebuild the event occurrence sequence with high accuracy from the observations and identify the crucial nodes on the attack paths.
\end{abstract}


\section{Introduction}

\par Due to the advances of trigger-action platforms (e.g. IFTTT \cite{IFTTT2020}) in IoT domain, IoT networks become more vulnerable towards malicious event injection attacks. Since IoT devices create a chain of interactions maintaining functional dependencies between entities and actions \cite{Celik2019b} \cite{Alam2021}, it is possible for adversaries to remotely inject malicious events somewhere in the interaction chain using a ghost device and activate a critical action through the exploitation of autonomous trigger-action scenario. For instance, an adversary can inject a fake thermometer reading of 110\textdegree F into the chain to initiate a critical \textit{window opening} action. 

\par There are a number of research efforts in the existing literature that attempt to solve the vulnerabilities caused by the trigger-actions in an IoT network. Most of them are designed to validate security properties by identifying the unsafe or insecure state transitions in the network \cite{Celik2019b} \cite{Nguyen-IoTSan-2018} \cite{Leonardo2018}. There is another line of research attempts where policy violations are addressed by checking sensitive user actions that may violate security policies \cite{Leonardo2018}. The research that is closest to our proposition is PEEVES \cite{sbirnbach2019}, where physical fingerprints of the devices are extracted using machine learning techniques to verify whether or not a certain event actually occurs.   

\par In this paper, we propose IoTMonitor, a security system that adopts a Hidden Markov Model based approach to determine the optimal attack path an attacker may follow to implement a trigger-action based attack, thus providing suggestions for subsequent patching and security measures. Our system examines the physical changes happening in an IoT environment due to the event occurrences, discovers the probabilistic relation between physical evidence and underlying events using the Baum-Welch algorithm \cite{Baum1967} \cite{Baum1968}, and discerns the optimal attack path using the Viterbi algorithm \cite{viterbi1967}. When the optimal attack path is determined, IoTMonitor identifies the crucial nodes in the path that the attacker must compromise to carry out the attack. Such information can be used for prioritizing security measures for IoT platforms.   

\par The contributions of the paper can be summarized as follows:
\vspace{-5pt}
\begin{itemize}
    \item We propose IoTMonitor, a Hidden Markov model based system that identifies the optimal attack path in a trigger-action IoT environment based on the probabilistic relation between actual IoT events and corresponding physical evidence;
    
    \item We implement the Baum-Welch algorithm to estimate transition and emission probabilities, and Viterbi algorithm to discern the attack path; 
    
    \item We propose an algorithm to detect the crucial nodes in an extracted optimal attack path, thus providing guidelines for subsequent security measures;
    
    \item We thoroughly evaluate the performance of IoTMonitor in detecting the optimal attack path and achieve high accuracy scores. 
\end{itemize}

\vspace{-5pt}
\par The rest of the paper is organized into four sections. In Section II, we define the attack landscape, discuss an attack scenario, and present the threat model. In Section III, we present IoTMonitor and discuss each component of it in detail. Later in Section IV, we present the evaluation results of our approach. Finally, in Section V, we conclude the paper by summarizing the methodology and outputs of our experiments, and presenting future extensions.  

\section{Attack Landscape}

\subsection{A Sample Attack Scenario}
\par Assume that Alice has a limited number of trigger-action enabled IoT devices including Smart Lock, Motion Detector, Accelerometer, Smart Light, Coffee Machine, and Smart Window. Alice controls each device through a mobile application from her cell phone. The devices communicate with each other through a hub. Since the platform supports trigger-action functionality, a device has the capability to trigger an action of another device.  

\par Alice sets up the trigger events as follows. When she unlocks the smart lock of the front door and walks in, the motion sensor in the living room detects the motion and activates ``home-mode''. The home-mode activation event automatically turns on the smart light. When the light is turned on, the events associated with coffee
grinding and window opening are triggered. 
When coffee is ready, Alice takes the coffee and enters into her bedroom by opening and subsequently closing the door. The vibration generated by the opening and closing operations of the door is measured by an accelerometer. Thus, a chain of events are triggered by the initial action. 

\par Now, Bob, an attacker, wants to compromise the smart window remotely when Alice is not at home and the front door is locked. His objective is to inject malicious events into the network to create a chain of interactions that eventually trigger the events associated with the window. 


\subsection{Threat Model}
\par We assume that the attacker knows the locations of the IoT devices in the target system but he does not have physical access to the home. He can eavesdrop on wireless communication taking place between devices and the hub. His goal is to perform a trigger-action attack by injecting fake events into the IoT network through ghost applications. The ghost applications impersonate target devices just by mimicking their characteristics and functionalities. Therefore, he does not need to deploy any real IoT devices to conduct the attack. 

\section{The IoTMonitor System}
\par Since the attacker exploits trigger-action functionality of IoT network to generate a chain of interactions by injecting fake events, we can thwart a trigger-action attack effectively if we can identify the optimal attack path the attacker may follow and perform security hardening on the crucial nodes in the attack path. In this research work, we propose \textit{IoTMonitor}, a system that discerns the optimal attack paths by analyzing physical evidence generated during the attack cycle, which are probabilistically correlated to the actual underlying events. IoTMonitor formulates the attack as a Hidden Markov Model (HMM) problem and solves it to determine the most likely sequence of events occur during an attack cycle and further identifies the crucial nodes in that sequence. Hence, in this paper, a \textit{node} represents an event occurring at a particular device. 



\subsection{Our Assumption} 
\par We assume that a configured trigger-action sequence contains $N$ events: $d_1, d_2,..., d_N$. The attacker injects fake events $\{d_i\}$ in the chain to achieve his final goal. Note that the attacker does not necessarily have to inject $d_1$ since he can wait for the occurrence of some real events to trigger the automatic chain occurrence of the rest of the events required to implement the attack. When an event is triggered, it causes some physical changes in the environment, which can be perceived as corresponding physical evidence $\{ph_i\}$ captured by an array of sensors and harnessed to verify the occurrence of that specific event. Note that some event may trigger non observable evidence, but others may trigger more than one evidence.    

\par Given this assumption, IoTMonitor models the trigger action scenario as a HMM problem, where physical evidence are visible to the analysis agent, but the actual events remain hidden. The tasks of the agent are to determine the probabilistic relation between events and evidence, and employ it to figure out the optimal attack path and diagnose the crucial nodes in that path.

\begin{figure}
    \centering
    \includegraphics[scale=0.35]{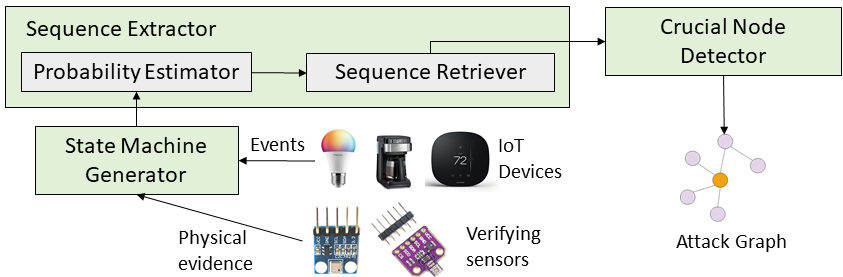}
    \caption{IoTMonitor System}
    \label{fig:IoTMonitor_system}
    \vspace{-10pt}
\end{figure}

\subsection{IoTMonitor}
The proposed IoTMonitor comprises three main components: 1) state machine generator, 2) sequence extractor, and 3) crucial node detector. Fig \ref{fig:IoTMonitor_system} shows the architecture of IoTMonitor. We discuss the components below in detail.

\subsubsection{\textbf{State Machine Generator}}
\par When events are triggered in the environment and the deployed sensors capture corresponding evidence per event occurrence, this component will construct a \textit{state machine} to represent how state changes in the environment due to the exploitation of trigger-action functionalities across a series of time instances $t=1, 2,..., T$. Hence, \textit{states} delineate useful information regarding the occurrence of different events $d_i$ and corresponding evidence $\{ph_i\}$. 

\par The state machine accommodates two types of states: 1) \textit{true states}, which correspond to the actual event occurrences, and 2) \textit{observation states}, which represent the physical evidence. Hence, the true states remain hidden, but the analysis agent leverages the observation states to infer the hidden true state sequence.


We define our state space as follows:
\begin{itemize}
    \item true state, $x_i$ : state responding to the occurrence of $d_i$
    \item observation state, $y_j$ : a subset of the physical evidence $\{ph_1, ph_2,..., ph_M\}$, which are emitted when the environment makes transition to a new state
\end{itemize}

\begin{figure}
    \centering
    \includegraphics[scale=0.35]{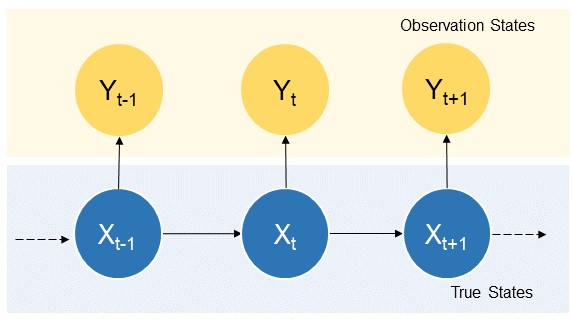}
    \caption{A Sample State Machine}
    \label{fig:State_machine}
    \vspace{-10pt}
\end{figure}

\par Hence, we assume that there are $N$ true states $X=\{x_1, x_2,..., x_N\}$, and $T$ observation states $Y = \{y_1, y_2,..., y_T\}$ in the state machine, where $X_t$ and $Y_t$, respectively, denote the true state and observation state at time $t$. Here, each $y_j$ contains a subset of the physical evidence $\{ph_1, ph_2,..., ph_M\}$, where the total number of evidence is $M$. Note that each observation state $Y_t$ in our experiment is determined with the help of a \textit{sliding window} function, which is discussed in detail in Section IV.

\vspace{1pt}
\par When the environment is in $x_i$ at time instance $t$ and makes a transition to any $x_j \in X$ at time instance $t+1$, it changes its true state with a \textit{transition probability} $q_{ij} \in Q$, which can be defined as:    
\vspace{-5pt}
\small
\begin{equation}
    \label{define-state-transition-probability}
        q_{ij} = Pr (X_{t+1}=x_j | X_t=x_i), \hspace{0.3cm}    1 \leq i,j \leq N
\end{equation}
\normalsize


Suppose, because of this state transition, the environment emits a new observation $y_k \in Y$ with an \textit{emission probability} $\mu_j(y_k) \in E$, which can be defined as:
\vspace{-2pt}
\small
\begin{equation}
    \label{define-emission-probablility}
    \begin{split}
        \mu_j(y_k) = Pr (Y_{t+1}=y_k | X_{t+1}=x_j), \hspace{0.3cm}   & 1 \leq j \leq N \\ & 1 \leq k \leq T
    \end{split}
\end{equation}
\normalsize

\par In the equation \eqref{define-state-transition-probability}, $Q = \{q_{ij}\}$ is termed as \textit{state transition probability distribution}, while $E = \{\mu_j(y_k)\}$ in the equation \eqref{define-emission-probablility} is termed as \textit{emission probability distribution}.

\par To model the attack as HMM, we need to generate an \textit{initial state distribution} $\sigma = \{\sigma_i\}$, such as:

\vspace{-3pt}
\small
\begin{equation}
    \begin{aligned}
        \label{initial-state-probability}
        \sigma_i = Pr(X_1 = x_i), \hspace{0.3cm} 1 \leq i \leq N    
    \end{aligned}
\end{equation}
\normalsize

Hence, $\sigma_i$ is the initial state distribution at time instance $t=1$.

\par Combining all the five aforementioned tuples, IoTMonitor models the trigger-action attack as an HMM problem $\big \langle N, M, Q, E, \sigma \big \rangle$ and solves it to determine the optimal attack path given a sequence of observation states. IoTMonitor also creates a parameter $\theta = (\sigma, Q, E)$, which is called the \textit{current model} of HMM.  

Figure \ref{fig:State_machine} shows a sample state machine where \textit{blue} circles represent the true states and \textit{yellow} circles represent the observation states. 


\textbf{Note:} For the rest of the paper, we call \textit{observation state} as only \textit{observation} sometimes and use the terms \textit{true state} and \textit{state} interchangeably to mean the same thing.

\subsubsection{\textbf{Sequence Extractor}}
\par Once the trigger action sequence is modeled as an HMM problem, IoTMonitor attempts to estimate the probability values and retrieve the optimal hidden state sequence from the observations. First, it starts with estimating the converged state distributions, transmission probabilities, and emission probabilities. Then, it seeks to figure out the underlying state sequence that maximizes the probability of getting a certain observation sequence. To accomplish both tasks, the \textit{sequence extractor} employs the following two subcomponents: a) probability estimator, and b) sequence retriever. The details of both subcomponents are described below.

\par a) \textbf{Probability Estimator}: Given a complete observation sequence $\langle Y_1, Y_2,..., Y_T \rangle$, the goal of this component is to determine the following: 
\vspace{-2pt}
\small
\begin{equation}
    \begin{split}
        \label{eqn1-baum-welch-general}
        \theta^* & = \underset{\theta}{argmax} \ Pr(Y_1, Y_2,..., Y_T | \theta)
    \end{split}
\end{equation}
\normalsize
\vspace{-20pt}


\par We use the Baum-Welch algorithm \cite{Baum1967} \cite{Baum1968} to iteratively update the current model $\theta$ and solve equation \eqref{eqn1-baum-welch-general}. It uses a \textit{forward-backward procedure} to find the maximum likelihood estimate of $\theta$ given a certain set of observations. We assume that each observation $Y_t$ is emitted by the environment at one discrete time instance $t=1,2,...,T$.

\vspace{3pt}
\par \textbf{Forward-backward Procedure}: Let $\alpha_t(i)$ and $\beta_t(i)$ are the probabilities of getting the observation sequences $ \langle Y_1, Y_2,..., Y_t \rangle$ and $\langle Y_{t+1}, Y_{t+2},..., Y_T \rangle$, respectively, while the system is being in the true state $x_i$ at time $t$.
So,
\vspace{-4pt}
\small
\begin{equation}
    \begin{aligned}
        \label{forward-backward-procedure}
        \alpha_t(i) &= Pr(Y_1, Y_2, ..., Y_t, X_t = x_i | \theta) \\
        \beta_t(i) &= Pr(Y_{t+1}, Y_{t+2}, ..., Y_T | X_t = x_i, \theta)  
    \end{aligned}
\end{equation}
\normalsize

\par We can compute $\alpha_t(i)$ and $\beta_t(i)$ using the following steps:

1. Initialization
\small
\begin{equation}
    \begin{aligned}
        \label{forward-backward-procedure-initialization}
        \alpha_1(i) &= \sigma_i \mu_i(y_1), \hspace{0.3cm} 1 \leq i \leq N \\
        \beta_T(i) &= 1, \hspace{0.3cm} 1 \leq i \leq N  
    \end{aligned}
\end{equation}
\normalsize
\vspace{-5pt}

2. Induction
\vspace{-10pt}
\small
\begin{equation}
    \begin{aligned}
        \label{forward-backward-procedure-induction}
        & \alpha_{t+1}(j) = \mu_j(y_{t+1}) \sum_{i=1}^N \alpha_t(i) q_{ij}, \hspace{0.2cm} 1 \leq t \leq T-1, \hspace{0.2cm} 1 \leq j \leq N  \\
        & \beta_t(i) = \sum_{j=1}^N q_{ij} \mu_j(y_{t+1}) \beta_{t+1}(j), \hspace{0.1cm} t = T-1, ...,2, 1, \hspace{0.2cm} 1 \leq i \leq N \\  
    \end{aligned}
\end{equation}
\normalsize

\par These two steps combined is called the \textit{forward-backward procedure}, and $\alpha_t(i)$ and $\beta_t(i)$ are termed as \textit{forward variable} and \textit{backward variable}, respectively.

\par Now, suppose $\delta_t(i)$ is the probability of the system being in the true state $x_i$ at time instance $t$ given the complete observation sequence $\langle Y_1, Y_2,..., Y_T \rangle$ and the current model $\theta$. We can define this probability in terms of the forward and backward variables $\alpha_t(i)$ and $\beta_t(i)$, i.e.,
\small
\begin{equation}
    \label{eqn1-update-delta}
    \begin{split}
        \delta_t(i) & = Pr(X_t = x_i | Y_1, Y_2, ..., Y_T, \theta) \\
                        & = \frac{Pr(X_t = x_i, Y_1, Y_2, ..., Y_T | \theta)}{Pr(Y_1, Y_2, ..., Y_T | \theta)} \\
                        & = \frac{\alpha_t(i)\beta_t(i)}{\sum_{j=1}^N \alpha_t(j)\beta_t(j)}
    \end{split}
\end{equation}
\normalsize

\par Again, given the complete observation sequence $\langle Y_1, Y_2,..., Y_T \rangle$ and the current model $\theta$, suppose, $\xi_t(i,j)$ is the probability of the system being in the true states $x_i$ and $x_j$ at time instances $t$ and $t+1$, respectively. So, 
\vspace{-3pt}
\small
\begin{equation}
    \label{eqn2-update-xi}
    \begin{split}
          \xi_t(i,j) & = Pr(X_t = x_i, X_{t+1} = x_j | Y_1, Y_2, ..., Y_T, \theta) \\
                        & = \frac{Pr(X_t = x_i, X_{t+1} = x_j, Y_1, Y_2, ..., Y_T | \theta)}{Pr(Y_1, Y_2, ..., Y_T | \theta)} \\
                        & = \frac{\alpha_t(i) q_{ij} \beta_{t+1}(j) \mu_j(y_{t+1})}{\sum_{i=1}^N \sum_{j=1}^N \alpha_t(i) q_{ij} \beta_{t+1}(j) \mu_j(y_{t+1})} 
    \end{split}
\end{equation}
\normalsize

\par Now, we can update the initial state distribution $\Bar{\sigma}_i$, transition probability $\Bar{q}_{ij}$, and emission probability $\Bar{\mu}_j(y_k)$ using these two parameters $\delta_t(i)$ and $\xi_t(i,j)$. The state distribution can be updated as:   
\small
\begin{equation}
    \label{eqn1-update-final-state-distribution}
        \Bar{\sigma}_i = \delta_1(i)
        \vspace{-3pt}
\end{equation}
\normalsize
\par where, $\delta_1(i)$ is the expected number of times the system is in the true state $x_i$ at time instance $t=1$.




\par To update the transition probabilities, we have to compute the ratio of \textit{the expected number of state transitions from $x_i$ to only $x_j$} (the numerator of the equation \eqref{eqn2-update-final-transition-probabalities}) and \textit{the expected number of transitions from $x_i$ to all other true states} (the denominator of the equation \eqref{eqn2-update-final-transition-probabalities}).

\small
\begin{equation}
    \label{eqn2-update-final-transition-probabalities}
        \Bar{q}_{ij} = \frac{\sum_{t=1}^{T-1} \xi_t(i,j)}{\sum_{t=1}^{T-1} \delta_t(i)}
\end{equation}
\normalsize

\par And to update the emission probabilities, we have to take the ratio of two other quantities: \textit{the expected number of times being in state $x_j$ and observing the observation $y_k$} (the numerator of the equation \eqref{eqn3-update-final-emission probabilities}), and \textit{the expected number of times being in state} $x_j$ (the denominator of the equation \eqref{eqn3-update-final-emission probabilities}).

\small
\vspace{-5pt}
\begin{equation}
    \label{eqn3-update-final-emission probabilities}
        \Bar{\mu}_j(k) = \frac{\sum_{t=1}^{T} 1_{(Y_t = y_k)} \delta_t(j)}{\sum_{t=1}^{T} \delta_t(j)}
\end{equation}
\normalsize

where, 
\small
\begin{equation}
    \label{eqn2-getting-observation_y_k}
        1_{(Y_t = y_k)} =  \left\{
        \begin{array}{@{}ll@{}}
            1, \hspace{0.5cm} \text{if } \hspace{0.1cm} Y_t = y_k \\
            0, \hspace{0.5cm} \text{Otherwise} \\
        \end{array}\right.
\end{equation}
\normalsize

\par The updated parameters $\Bar{\sigma} = \{\Bar{\sigma}_i\}$, $\Bar{Q} =  \{ \Bar{q}_{ij} \}$, and $\Bar{E} = \{ \Bar{\mu}_j(y_k) \}$ now constitute the new model $\Bar{\theta} = (\Bar{\sigma}, \Bar{Q}, \Bar{E})$. We need to iterate the equations \eqref{eqn1-update-final-state-distribution} \eqref{eqn2-update-final-transition-probabalities}, and \eqref{eqn3-update-final-emission probabilities} until we find $\Bar{\theta} \approx \theta$. This convergence is guaranteed in \cite{Baum1968} by Baum et al., where it is ensured that either 1) the initial model $\theta$ defines a critical point in the likelihood function where $\Bar{\theta}=\theta$, or 2) $\Bar{\theta}$ explains the observation sequence $\langle Y_1, Y_2,..., Y_T \rangle$ more suitably than $\theta$, i.e. $Pr(Y_1, Y_2,..., Y_T | \Bar{\theta}) > Pr(Y_1, Y_2,..., Y_T | \theta)$ \cite{Rabiner1989}. 
\BlankLine

\par b) \textbf{Sequence Retriever}: Once the probability estimator determines the converged HMM model $\theta^*$, now, it is job for the \textit{Sequence Retriever} to extract the optimal sequence of hidden events using Viterbi algorithm \cite{viterbi1967}. Given a particular observation sequence $\langle Y_1, Y_2,..., Y_t \rangle$ at time instance $t$ and $Y_t = y_k$, the goal here is to determine the following: 
\small
\begin{equation}
    \label{viterbi-objective-eqn}
    \begin{split}
         \omega _t(i) &= \underset{x_1,..., x_{i-1}}{max} \Big \{ Pr(X_1 = x_1,...,X_t = x_i, Y_1,..., Y_t = y_k| \theta) \Big \} \\
        & = \underset{x_1, x_2,..., x_{i-2}}{max} \bigg \{ \underset{x_{i-1}}{max} \Big \{\omega _{t-1}(i-1) q_{(i-1)(i)} \Big \} \mu_t (y_k) \bigg \}, \\ & \hspace{4.5cm} 2 \leq t \leq T, \hspace{0.1cm} 1 \leq i \leq N
    \end{split}
\end{equation}
\normalsize

Hence, $\omega _t(i)$ represents the maximum probability of the occurrence of a particular state sequence $\langle x_1, x_2,..., x_i \rangle$ at time instance $t$ that corresponds to the aforementioned observation sequence $\langle Y_1, Y_2,..., Y_t \rangle$.

\par The equation \eqref{viterbi-objective-eqn} can be solved recursively to determine the highest probability of the occurrence of a complete state sequence $\langle x_1, x_2,..., x_N \rangle$ for the time instance $2 \leq t \leq T$ given that $\omega _1(i) = \sigma_i \mu_i(y_1)$. The recursion stops after computing $\omega _T(i)$ such as:
\small
\begin{equation}
    \label{viterbi-omegha_final_instant}
    \omega _T^* = \underset{1 \leq i \leq N}{max} \ \omega _T(i)
\end{equation}
\normalsize

\par But to obtain the optimal hidden sequence, we must trace the arguments that maximize the equation (\ref{viterbi-objective-eqn}) during each recursion. To achieve that, we introduce a variable $\chi$ to hold all the traces such as:  
\small
\begin{equation}
    \label{viterbi-trace-arguments}
    \chi _t(i) = \underset{1 \leq i \leq N }{argmax} \Big \{ \omega _{t-1}(i-1) q_{(i-1)(i)} \Big \}, 2 \leq t \leq T, 1\leq i \leq N
\end{equation}
\normalsize
Note that $\chi _1(i) = 0$ for $t=1$ because we start tracing the states for the very first time at time instance $t=2$ once we have at least one previous state.



\par Once we have $\chi _T(i)$, all we need is backtracking through the traces to discern the optimal hidden sequence such as:
\small
\begin{equation}
    \label{viterbi-backtracking}
        \psi _t ^ * = \chi_{t+1} (\psi_{t+1} ^ *), \ t = T-1, ....., 2, 1   
\end{equation}
\normalsize

Hence, $\psi _T ^ * (i)  = \chi _T(i)$, and $\Upsilon = \{\psi_1 ^*, \psi_2 ^*,..., \psi_T ^*\}$ is the extracted optimal sequence. Note that each $\psi_t ^* \in \Upsilon$ represents a true state in $X$.

\vspace{5pt}
\normalsize
\subsubsection{\textbf{Crucial Node Detector}}
\par After the \textit{sequence retriever} extracts the hidden optimal sequence $\Upsilon = \{\psi_1 ^*, \psi_2 ^*,..., \psi_T ^* \}$, the component \textit{crucial node detector} applies Algorithm \ref{algo:crucial_node_detection} to detect the crucial events in the attack chain the attacker must compromise to successfully implement the attack. Hence, the most frequently triggered events are defined as \textit{crucial events}. 

\par If there are $p$ number of different extracted sequences $\Upsilon_1, \Upsilon_2,..., \Upsilon_p$ for $p$ different attempts, the Algorithm \ref{algo:crucial_node_detection} first determines the \textit{longest common subsequence} $S_i$ between each $\Upsilon_i$ and the original sequence $X = \{x_1, x_2, ..., x_N\}$. Later, it computes the \textit{SCORE} value for each pair of states in the subsequence such as:
\small
\begin{equation}
    \label{equation:SCORE_definition}
    \begin{split}
        & SCORE \Big[S_i[j],S_i[j+1] \Big] =  \text{number of times a pair} \\
        & \big \{ S_i[j],S_i[j+1] \big \} \text{ is present in the subsequence} \    
    \end{split}
\end{equation}
\normalsize
\vspace{-5pt}

\vspace{-5pt}
\begin{algorithm}[h]
\footnotesize
\caption{Crucial node detection algorithm}
\label{algo:crucial_node_detection}
\KwIn{$X, \Upsilon_1, \Upsilon_2, ...., \Upsilon_p$}
\KwOut{ Pairs of true states responding to the most frequently triggered events}
\BlankLine
\begin{algorithmic}[1]
    \STATE $i \gets 1$
    \WHILE{$i \leq p$}
        \STATE $S_i \gets$ LCS between $X$ and $\Upsilon_i$ \tcp{\footnotesize LCS = Longest Common Subsequence}
        \FOR{$j \gets 1$ \KwTo $(|S_i|-1)$}
            \STATE $E[i, j] \gets \{S_i[j], S_i[j+1]\}$ \\
            \IF{$E[i, j]$ not in $SCORE.Keys()$} 
                \STATE $SCORE[E[i,j]] \gets 1$ \\
            \ELSE
                \STATE $SCORE[E[i,j]] \gets SCORE[E[i,j]] + 1$ 
            \ENDIF
        \ENDFOR
    \ENDWHILE
            
    \RETURN $\underset{E[i,j]}{argmax} \ (SCORE[E[i,j]])$
\end{algorithmic}
\end{algorithm}
\vspace{-5pt}

\par Finally, the algorithm updates the \textit{SCORE} values based on the presence of pairs in all subsequences and retrieves the pairs with the maximum \textit{SCORE} value. It may output a number of pairs of states, such as $\{ x_{c_i}, x_{c_j} \}$, where there is a crucial state transition in the state machine from $x_{c_i}$ to $x_{c_j}$. Our goal is to identify the events (we call them \textit{nodes}) associated with such transitions that are exploited by the attackers to compromise the chain.

\vspace{5pt}
\par \textbf{A Simple Example}
\par Suppose, there is a sequence of states (responding to some triggered events): \{\textbf{door-opened, light-on, camera-on, fan-on, window-opened}\}. And after making three separate attempts, the sequence retriever returns the following three sequences:

Sequence-1: \{door-opened, light-on, light-on, camera-on, fan-on\}, Sequence-2: \{fan-on, light-on, camera-on, fan-on, window-opened\}, Sequence-3: \{door-opened, light-on, camera-on, window-opened, fan-on\}.

\par Now, if we apply Algorithm \ref{algo:crucial_node_detection} on this scenario, we find that the pair \{\textbf{light-on, camera-on}\} obtains the highest score. Consequently, we can conclude that the transition from the state \textbf{light-on} to \textbf{camera-on} is the most vital one in the state machine, and the nodes associated with those states are the most crucial ones in the chain. IoTMonitor identifies these crucial nodes so that we can perform security hardening to minimize the attacker's chance of compromising an IoT network. The security hardening part is out of the scope of this paper, and we plan to incorporate such capability in the extended version of IoTMonitor in recent future.   





\section{Results and Evaluation}
\par To evaluate the performance of IoTMonitor, we utilize the PEEVES dataset \cite{sbirnbach2019} that records IoT event occurrences from 12 different IoT devices and sensors measurements from 48 deployed sensors to verify those events. We use 24-hours data for our experiment, and our experiment executes on a 16 GB RAM and 4 CPU core system.    

\subsection{Dataset Processing}
\par Our experiment mainly deals with three types of data: 1) event data (used as true states) 2) sensor measurements (used as observations), and 3) timestamps. We concentrate only on those event occurrences which can be verified by the sensor measurements. Since sensor measurements here capture the physical changes that have happened in the environment due to the event occurrences, they can be used to crosscheck whether a certain event has occurred. We conceptualize the function \textit{sliding window} to determine whether an event is verifiable. Hence, the function provides us with a time window (in milliseconds) $w_i$ that starts at the timestamp of a particular event occurrence. After an event occurrence is recorded at time instance $t_i$, if we find the necessary sensor measurements to verify that occurrence within the time instance $t_i + w_i$, we consider that event verifiable and keep it in the sequence of events occurred. Otherwise, we discard it from the sequence. In our experiment, we consider 20 such sliding windows with the size between 105 milliseconds and 200 milliseconds with an increase of 5 milliseconds. 

\subsection{Experiment Setting}
\par At the beginning of our experiment, we choose Gaussian distribution to randomly assign the transition probabilities and initial state probabilities for each true state. On the other hand, we use Dirichlet distribution to assign the emission probabilities. We use the same seed value for each execution.   

\subsection{Probability Estimation Time} 
\par
\begin{figure}[h]
    \centering
    \includegraphics[scale=0.33]{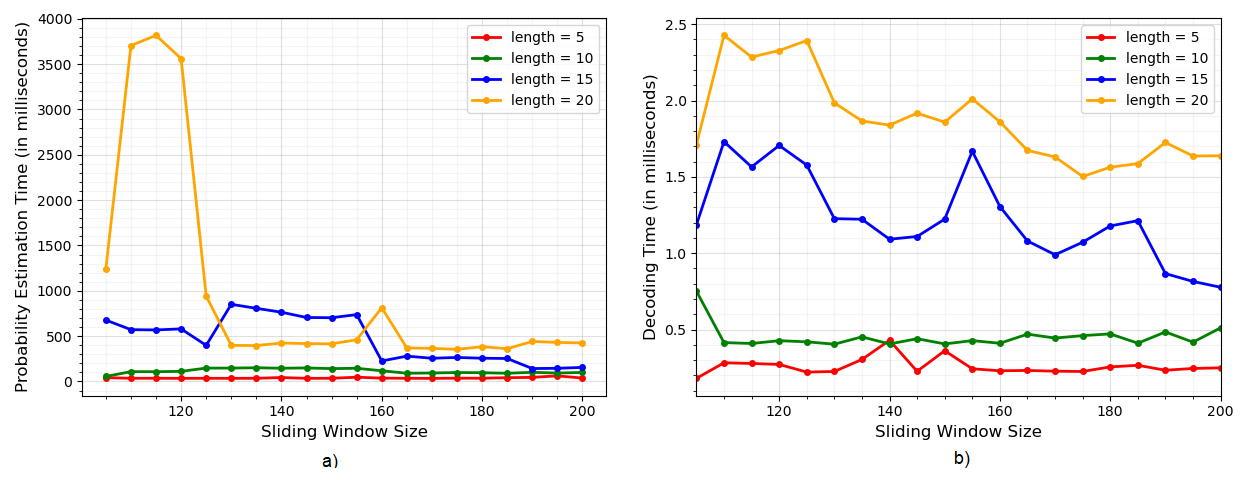}
    \vspace{-10pt}
    \caption{a) Probability estimation time with respect to sliding window size and length of the event sequence; b) Decoding time with respect to sliding window size and length of the event sequence }
    \label{fig:estimation_time_decoding_time}
    \vspace{-15pt}
\end{figure}

 Probability estimation time represents the time required to estimate the converged transition probability distribution $Q$ and the emission probability distribution $E$. Figure \ref{fig:estimation_time_decoding_time}(a) presents the estimation time for four different sequences of events of different lengths (5, 10, 15, and 20) against a range of sliding windows. In the figure we show the average estimation time after 10 executions.    

\par As we can see from Figure \ref{fig:estimation_time_decoding_time}(a), the longest estimation time is $<4$ seconds for the sequence length of 20, while in most cases, it is $<0.5$ seconds. As the window size increases, the estimation time starts to decrease and stabilize. There are a few exceptional cases where the estimation time increases sharply for a increase in window size. For example, when the window size increases from 105 to 110 for the sequence of length 20, we see a sudden spike. We examine the source data and find that this spike is caused by the appearance of two new events that were not present earlier. Since the number of unique events increases and repetition of same events decreases in the sequence, the initial state distribution and transition probabilities are needed to be adjusted which costs adversely to the total estimation time. However, this type of exception is transient, and the graph stabilizes eventually. We do not present the estimation time for the sequences of lengths $>20$ in the Figure \ref{fig:estimation_time_decoding_time}(a) since we observe very little change in pattern for those sequences.          

\vspace{-8pt}



\subsection{Decoding Time}
\par Decoding time represents the time required to extract the hidden sequence when we have the converged $\theta^*$. Similar to probability estimation time, we take average decoding time after 10 executions. The Figure \ref{fig:estimation_time_decoding_time}(b) presents the decoding time for four different sequences of events with lengths 5, 10, 15, and 20 against a range of sliding windows.

\par If we look at the graph at Figure \ref{fig:estimation_time_decoding_time}(b), we see that the decoding time decreases when the window size increases. The longest decoding time we get is $<2.5$ milliseconds which is very fast for the retrieval of hidden event sequences. Although we see few little temporary spikes for the length 15 after sliding window 150, we still achieve $<2.0$ milliseconds as the decoding time.    

\begin{figure}[h]
    \centering
    \includegraphics[scale=0.40]{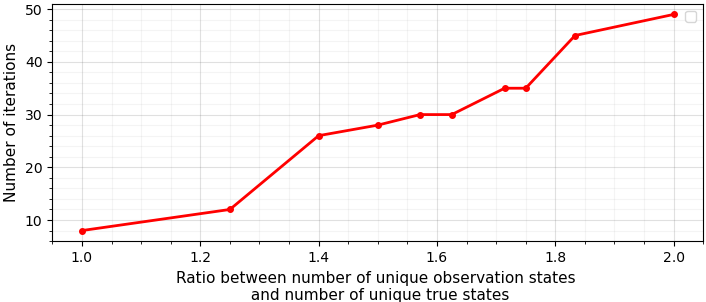}
    \caption{Number of iterations to estimate the converged transition probabilities and emission probabilities with respect to the ratio between number of observation states and number of true states}
    \label{fig:computational overhead}
    \vspace{-10pt}
\end{figure}

\vspace{-3pt}
\subsection{Computational Overhead}
\par Since our experiment dedicates most of the computation time to estimate the probabilities, we measure \textit{computational overhead} as the total number of iterations of the \textit{forward-backward procedure} required to reach the convergence of transition probabilities and emission probabilities. In Figure \ref{fig:computational overhead}, we present the required total number of iterations (in y-axis) with respect to the ratio between \textit{the total number of unique observation states} and \textit{the total number of unique true states} (in x-axis). We can see that, the computational overhead increases roughly linearly with the ratio.   

\begin{figure}[h]
    \centering
    \includegraphics[scale=0.32]{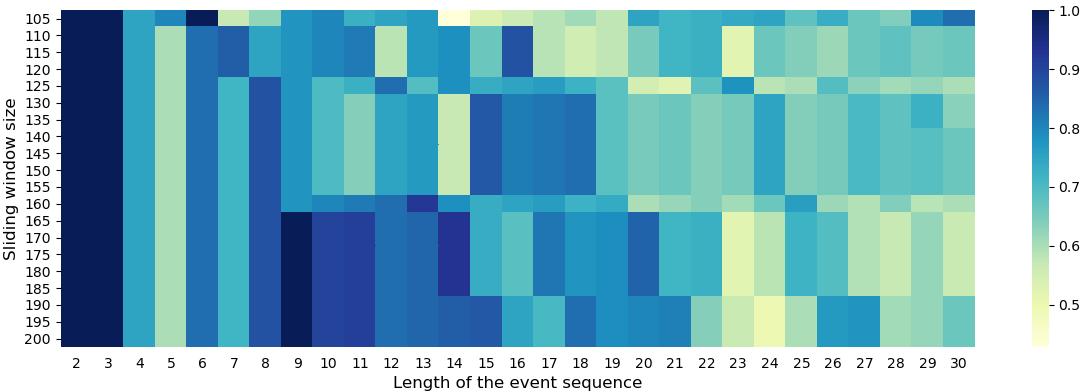}
    \vspace{-15pt}
    \caption{Accuracy score vs Sliding window size vs Length of the event sequence}
    \label{fig:accuracy_score}
    \vspace{-5pt}
\end{figure}



\subsection{Accuracy Score}
\par To determine how accurately the extracted hidden sequence of events represent the real events, we compute f-score for 29 different sequence of events starting with the length 2 and ending at length 30. We do not consider the sequence with length 1 because it does not offer any uncertainty in terms of transition and emission probability. We present a heatmap to visually show the correlation among accuracy score, sliding window size and length of the event sequence. In Figure \ref{fig:accuracy_score}, the accuracy scores are presented as colors. 

\par As we can see, when the length of event sequence is $<15$, the increase in window size after 160 assures a very high accuracy score. We even get the accuracy score of 1.0 in some occasions. There is only one exception for the sequence of length 5. We see a decrease in accuracy score after the window size 105, and that's because we see a completely new sequence for the window sizes 110 to 200. Similar pattern also arises, although to a less extent, for the sequence of length 7. But it is quite evident that the increase in window size for the smaller lengths ensures higher accuracy score (equals or close to 1.0). When the length increases to a considerable extent, we start to see the impact of sliding windows on the accuracy score diminishing slowly. Since our system emphasizes on the functional dependencies (in terms of transition probability) of the events to extract the hidden  sequence, the longer the sequence becomes, the looser are the dependencies. 


\vspace{-5pt}
\section{Conclusion}
\par In this research work, we propose IoTMonitor that focuses on the extraction of the underlying event sequence using HMM approach given a set of physical evidence emitted during a trigger-action based attack in an IoT environment. We use the Baum Welch algorithm to estimate transition and emission probabilities, and Viterbi algorithm to extract the underlying event sequence. 
Our experiments show that both probability estimation and sequence extraction operations converge reasonably fast.  
In terms of accuracy score, IoTMonitor achieves 100\% in multiple cases and $\geq 90\%$ in a number of cases. We draw a heatmap to visually show the correlation among accuracy score, sliding windows, and length of the event sequences. We also present an algorithm to identify the crucial events in the extracted sequence which the attackers wish to compromise to implement a trigger-action attack.  

Immediate extensions to our approach include the following efforts. First, we currently focus on the crucial nodes that appear in multiple attack paths. If we extend our research to an attack graph, we can identify crucial node pairs on different attack paths. Second, the physical evidence collected by sensors could contain noises or even inaccurate data. We will improve our algorithm to provide more robust attack detection capability for IoT platforms.

\bibliographystyle{./bibliography/IEEEtran}
\bibliography{./bibliography/IEEEabrv,./bibliography/IEEEexample}


\end{document}